\documentclass[%
reprint,
 amsmath,amssymb,
 aps,
prb,
]{revtex4-1}

\usepackage{graphicx}
\usepackage{dcolumn}
\usepackage{bm}

\usepackage{mathtools}
\usepackage{color}
\definecolor{mygray}{gray}{0.6}

\begin{document}
\preprint{}

\title{Effects of Disorder on a 1-D Floquet Symmetry Protected Topological Phase}

\author{Yuval Gannot}
 \email{ygannot@berkeley.edu}
\affiliation{%
Department of Physics, University of California, Berkeley.
}%


\begin{abstract}

Periodically driven systems, also known as Floquet systems, can realize symmetry protected topological (SPT) phases that cannot be found in equilibrium.  
Here, we seek to understand the effects of strong disorder on such SPT phases, working with a one-dimensional Floquet system belonging to Altland-Zirnbauer class BDI. 
Using the transfer matrix method, we find that the disordered system hosts an array of trivial, topological, and Floquet topological phases. We then explore the phase diagram in the case of uniformly distributed bonds. Our analytic results are confirmed by a numeric computation of the bulk topological invariants. Although all states are generically localized, near phase transitions the localization length of topological edge states diverges. We derive the critical exponents governing these divergences, finding that they take on equilibrium values, despite the fact that we are considering phases unique to non-equilibrium.
We also discuss how our work sets the stage for studying interacting Floquet SPT phases.

\end{abstract}

\pacs{Valid PACS appear here}
\maketitle


\section{\label{intro}Introduction and Motivation}
Symmetry protected topological phases (SPT phases) have received significant attention over the past several years \cite{Hasan10,Qi10,Senthil15}. More recently, the notion of SPT phases was generalized to periodically driven systems \cite{Takashi09, Kitagawa10, Jiang11, Kitagawa11b, Lindner11, Rudner13, Asboth14,Nathan15}
, also known as Floquet systems, and closely related quantum walk systems \cite{Kitagawa10b, Kitagawa11, Asboth13}. 
This generalization includes phases that have no counterpart in static systems, so that these so-called Floquet SPT phases require entirely new theoretical classification. 
For instance, two-dimensional Floquet systems can have chiral edge modes even when all bands have zero Chern number \cite{Rudner13}, and one dimensional Floquet systems and quantum walks can have protected edge states despite having zero bulk winding number \cite{Kitagawa11}.

In the presence of interactions, a translationally invariant Floquet system continuously absorbs energy from the periodic drive and heats up to an infinite temperature \cite{DAlessio14, Lazarides14, Ponte15}, thus precluding the existence of an SPT phase or other types of quantum order \cite{Khemani15}. 
However, thermalization can be avoided if strong disorder is incorporated so that the system is many-body localized (MBL) \cite{Basko06,Nandkishore15}; this has been demonstrated in both static \cite{Oganesyan07, Pal10}
 and periodically driven \cite{Abanin14, Ponte14, Lazarides15} systems.
In this case, various types of quantum order have been shown to survive in the excited states of static systems \cite{Huse13, Pekker14, Chandran14,Kjall14,Vosk14,Bahri15}
and, more recently, in Floquet systems \cite{Khemani15}.
While the eventual goal along these lines is to understand SPT phases in strongly disordered, interacting Floquet systems, in this work we first tackle the aspect of disorder, analyzing the interplay between single-particle localization and the SPT phases of a one-dimensional noninteracting Floquet system. 

The model we study belongs to Altland-Zirbauer symmetry class BDI. Its nontrivial phases, which are protected by chiral symmetry, are characterized by edge states with so-called quasienergy equal to $0$ or $\pi/T$, where $T$ is the period of the Floquet drive. Our goal is to find the topological phase diagram in the presence of disorder, to confirm that the system is generically single-particle localized, and to characterize the transition between distinct topological phases. To analyze this system, we adapt two methods from the study of time-independent disordered systems: the transfer matrix method and the so-called real-space winding number topological invariant \cite{Mondragon-Shem14, Song14}.

\section{\label{floq_theory}Floquet Theory}
In this section we briefly review Floquet theory. A Floquet system has a Hamiltonian $H(t)$ that is periodic in time $t$ with period $T$. 
$H(t)$ admits solutions $|\psi(t)\rangle$ to the time-dependent Schrodinger equation that satisfy
$|\psi(t+T)\rangle = e^{-iET}|\psi(t)\rangle$. Here $E$ is a real parameter defined modulo $2\pi/T$ called the quasienergy. The solutions $|\psi(t)\rangle$ are called Floquet states \cite{Kitagawa10, Rudner13}. Equivalently, the Floquet states satisfy the Floquet eigenvalue equation 
\begin{equation}
U_{t+T,t}|\psi(t)\rangle = e^{-iET}|\psi(t)\rangle,
\label{eq:floquet}
\end{equation}
where $U_{t+T,t}$ is the operator that evolves states from time $t$ to time $t+T$. We will usually pick the initial time to be $0$, writing  $U_{T,0} = U_T$. We can also define $H_{\text{eff}} = \frac{i}{T} \log U_{T}$ so that 
\begin{equation}
H_{\text{eff}}|\psi(t)\rangle = E|\psi(t)\rangle.
\label{eq:heff}
\end{equation}
The interpretation of (\ref{eq:heff}) is that if the system is examined stroboscopically with period $T$, $H(t)$ simulates a time independent Hamiltonian $H_{\text{eff}}$ whose energies are defined modulo $2 \pi/T$ \cite{Kitagawa10, Rudner13}. 

The notion of an SPT phase generalizes naturally to Floquet systems. A nontrivial Floquet SPT phase is characterized by Floquet edge states robust to perturbations that preserve the symmetries of the effective Hamiltonian $H_{\text{eff}}$. The only subtle point is that we are interested in the symmetries of $H_{\text{eff}}$, not those of the instantaneous Hamiltonian $H(t)$ \cite{Kitagawa10}.

\section{\label{describe_ham}Hamiltonian and its Phases}
We now introduce the model that we study in this work, and then discuss its symmetries and its resulting topological phases. The model consists of spinless fermions hopping on a one-dimensional lattice containing two sites $A$ and $B$ per unit cell. It is obtained by adding disorder to a special case of the periodically driven Su-Schreiffer-Heeger model introduced in Ref. \onlinecite{Asboth14}.

Let $\bm{c}$ denote the vector of site annihilation operators:
\begin{equation}
\bm{c} =
\begin{pmatrix}
 c_{A1} \\
 c_{B1} \\
 \vdots \\
 c_{AN} \\
 c_{BN}
 \end{pmatrix} ,
 \end{equation}
 where $c_{A i}$ and $c_{B i}$ annihilate a fermion at site A and B, respectively, in unit cell $i$. The Hamiltonian is then defined by
 \begin{equation}
 \label{eq:many_body}
 H(t)=\bm{c}^{\dagger}\mathcal{H}(t)\bm{c}
 \end{equation}
  where the first-quantized matrix $\mathcal{H}(t)$ is given by
\begin{equation}
\mathcal{H}(t) = 
	\begin{cases}
	\mathcal{H}_1(\{J_i\}) & 0 \le t < \frac{1}{4}T \\
	\mathcal{H}_2(\{K_i\}) & \frac{1}{4}T \le t < \frac{3}{4}T \\
	\mathcal{H}_1(\{J_i\}) & \frac{3}{4}T \le t \le T 
	\end{cases}	
	\label{eq:ham}	
\end{equation}
with $\mathcal{H}_1$ and $\mathcal{H}_2$ Hermitian matrices satisfying
\begin{subequations}
\begin{align}
\bm{c}^{\dagger}\mathcal{H}_1(\{J_i\})\bm{c}  &= \sum_{1=i}^N 2J_{i} (c_{Ai}^{\dagger}c_{Bi} + c_{Bi}^{\dagger}c_{Ai}) \\
\bm{c}^{\dagger}\mathcal{H}_2(\{K_i\})\bm{c}  &= \sum_{1=i}^{N-1} 2K_{i} (c_{Ai+1}^{\dagger}c_{Bi} + c_{Bi}^{\dagger}c_{Ai+1}).
\end{align}
\end{subequations} 
Here $\{J_{i}\}$ and $\{K_{i}\}$ are two sets of independent, identically distributed random variables. The first-quantized time evolution operator is then given by
\begin{align}
\label{eq:explicit_UT}
\mathcal{U}_{T} & =\mathcal{T}e^{-i\int_0^T \mathcal{H}(t)} \nonumber \\ 
& = e^{-\frac{i}{4}T\mathcal{H}_1(\{J_i\}) }e^{-\frac{i}{2}T\mathcal{H}_2(\{K_i\})}e^{-\frac{i}{4}T \mathcal{H}_1(\{J_i\})},
\end{align}
where $\mathcal{T}$ is the time ordering symbol.

In the translationally invariant limit, variants of this model have been explored in the context of driven superconductors \cite{Jiang11} and 1-D quantum walks \cite{Kitagawa11, Asboth13}. The so-called ``simple quantum walk'', a special case of the quantum walk analogue of  (\ref{eq:many_body}), has been studied numerically in the presence of disorder \cite{Tarasinski14, Obuse11}.

We now discuss the symmetries of the system and its resulting topological classification. We consider particle-hole, time reversal, and chiral symmetries, as dictated by the``ten-fold way'' of classifying free fermion topological insulators according to Altland-Zirnbauer symmetry classes \cite{Ryu10}. The conditions for a system to respect these symmetries can be translated from conditions on $\mathcal{H}_{\text{eff}}$ to conditions on $\mathcal{U}_{T}$. The result is that particle-hole symmetry requires $\mathcal{P}\mathcal{U}_{T} \mathcal{P} =  \mathcal{U}_{T} $ for antiunitary $\mathcal{P}$, time reversal symmetry requires $\mathcal{T}\mathcal{U}_{T} \mathcal{T} =  \mathcal{U}_{T} ^{\dagger} $ for antiunitary $\mathcal{T}$, and chiral, or sublattice, symmetry requires $\mathcal{S}\mathcal{U}_{T} \mathcal{S} =  \mathcal{U}_{T} ^{\dagger} $ for unitary $\mathcal{S}$ \cite{Tarasinski14, Kitagawa10}.

Using $\mathcal{P}= (\sigma_z \otimes 1) K$, $\mathcal{T} = K$ and $\mathcal{S}= \sigma_z \otimes 1$, where $\sigma_z$ acts within a unit cell, $1$ acts between different unit cells and $K$ is the complex conjugation operator, we see that (\ref{eq:explicit_UT}) respects all three symmetries \cite{Asboth14}, and that all three operators square to $+1$. Therefore it belongs to symmetry class BDI \cite{Ryu10}.

Static systems in class BDI have a $\mathbb{Z}$ classification of topological phases \cite{Ryu10}, where $\nu \in \mathbb{Z}$ counts the number of zero energy edge states localized on sublattice $A$ minus the number of zero energy edge states localized on sublattice $B$. This classification relies crucially on the fact that $\mathcal{S}$ maps states with energy $E$ into states with energy $-E$, and that in static systems $E=0$ is the only energy satisfying $E= -E$. However, in driven systems quasienergy $E=\pi/T$ also satisfies $ E = -E$. Hence the topological classification becomes $\mathbb{Z} \times \mathbb{Z}$, with one integer $\nu_0$ counting edge states with $E=0$ and a second integer $\nu_{\pi}$ counting edge states with $E=\pi/T$. The phases with nonzero $\nu_{\pi}$ are unique to Floquet systems; it is impossible to continuously deform a Floquet system with nonzero $\nu_{\pi}$ into a time independent system without encountering a phase transition \cite{Kitagawa11}.

Throughout this work it will be instructive to compare our results with those for the analogous time independent Hamiltonian $\mathcal{H_{\text{ti}}}$ belonging to class BDI:
\begin{equation}
\label{eq:H_ti}
\mathcal{H_{\text{ti}}}= \mathcal{H}_1(\{J_i\}) +\mathcal{H}_2(\{K_i\}).
\end{equation}
$\mathcal{H_{\text{ti}}}$ is the Hamiltonian of the usual Su-Schrieffer-Heeger model. Up to an additional phase factor $i$ in the inter-cell hoppings breaking the symmetry down to class AIII, this is the model studied in the presence of strong disorder by Ref. \onlinecite{Mondragon-Shem14}. The phase factor $i$ does not affect the results that we will cite.

\section{\label{transfer_matrix} Transfer Matrix Method}
Here we discuss the the transfer matrix of the system and how it leads to a simple formula for $\nu_0$ and $\nu_{\pi}$. In general, the transfer matrix method is useful for solving 1-D lattice models with finite range hopping when translation symmetry is not available. First we review its application to static systems. In this case the goal is to convert a large (possibly infinite) eigenvalue equation $\mathcal{H}\psi = E\psi$ to a small matrix problem of the form
\begin{equation}
\begin{pmatrix}
\psi_{i+t} \\
\psi_{i+t+1} \\
\vdots \\
\psi_{i+t+s}
\end{pmatrix}
=
T_i(E)
\begin{pmatrix}
\psi_{i} \\
\psi_{i+1} \\
\vdots \\
\psi_{i+s}
\end{pmatrix}, 
\end{equation}
where $ 1 \le t \le s+1$. Numerically iterating the transfer matrix allows one to compute its Lyapunov exponents, the smallest in magnitude of which gives the inverse localization length of the system \cite{Kramer93}. When symmetries constrain the form of the transfer matrix it is possible to obtain analytic information about the eigenstates. 

For driven systems, we might hope to find a transfer matrix representation of the Floquet eigenvalue equations $\mathcal{U_T} \psi = e^{-iET} \psi$ or $\mathcal{H}_{\text{eff}}\psi = E\psi$. In general this is a hopeless task, because $\mathcal{U}_T$ and $\mathcal{H}_{\text{eff}}$ will have infinite (though exponentially decaying) range even if the instantaneous Hamiltonian has finite range. Hamiltonian (\ref{eq:ham}) is special in this respect, because it is a finite sequence of block diagonal Hamiltonians, so $\mathcal{U}_T$ has finite range. Indeed, following the steps of Appendix \ref{app:computation} we find that the transfer matrix has the two dimensional structure
\begin{equation}
\label{eq:T_form}
\begin{pmatrix}
\psi_{Ai+1} \\
\psi_{Bi+1}
\end{pmatrix}
= T_i(E)
\begin{pmatrix}
\psi_{Ai} \\
\psi_{Bi}
\end{pmatrix}.
\end{equation}
$T_i(E)$ is derived in Appendix \ref{app:computation} for all quasienergies. Here we only quote the results for $E=0$ and $\pi/T$, which will be used to study the topological phases of the system:
\begin{widetext}
\begin{subequations}
\begin{align}
T_i(0) = & 
\begin{pmatrix*}[c]
-\dfrac{\sin   \frac{1}{2} {J}_iT}{\cos \frac{1}{2} {J}_{i+1}T} \cot \frac{1}{2} {K}_iT  &  0\\
0 & -\dfrac{\cos \frac{1}{2} {J}_iT}{\sin  \frac{1}{2} {J}_{i+1}T} \tan \frac{1}{2} {K}_iT
\end{pmatrix*} \\
T_i(\pi/T) = & 
\begin{pmatrix*}[c]
\dfrac{\sin   \frac{1}{2} {J}_iT}{\cos \frac{1}{2} {J}_{i+1}T} \tan \frac{1}{2} {K}_iT  &  0\\
0 & \dfrac{\cos \frac{1}{2} {J}_iT}{\sin  \frac{1}{2} {J}_{i+1}T} \cot \frac{1}{2} {K}_iT
\end{pmatrix*}. 
\end{align}
\end{subequations} 
\end{widetext}
Note that $T_i(0)$ and $T_i(\pi/T)$ are both diagonal. Indeed, this follows from the form (\ref{eq:T_form}) and chiral symmetry: since edge states with quasienergy $0$ or $\pi/T$ can be chosen to be eigenstates of the chiral symmetry operator $\mathcal{S} = \sigma_z \otimes 1$, it follows that
\begin{equation}
\sigma_z
\begin{pmatrix}
\psi_{Ai+1} \\
\psi_{Bi+1}
\end{pmatrix}
= T_i(0)
\sigma_z
\begin{pmatrix}
\psi_{Ai} \\
\psi_{Bi}
\end{pmatrix},
\end{equation}
so that $\sigma_z T_i(0) \sigma_z = T_i(0)$. The same result holds for $E=\pi/T$, which implies that $T_i(0)$ and $T_i(\pi/T)$ are both diagonal.

To find if there exists an edge state at these energies and hence determine $\nu_0$ and $\nu_{\pi}$, we use a semi-infinite geometry extending from $i=1$ to $\infty$, and ask whether an initial wave function amplitude $( \psi_{A1},\psi_{B1})$ satisfying boundary conditions at the edge decays under repeated application of the transfer matrix. The proper boundary conditions are derived along with the transfer matrix in Appendix \ref{app:computation}, the result being $\psi_{B1} = 0$  and $\psi_{A1} = 0$ for $E=0$ and $\pi/T$, respectively. This shows immediately that this Hamiltonian only supports $\nu_0 = 0$, $1$, and $\nu_{\pi}=0$, $-1$. 

Note that these boundary conditions are independent of the disorder realization. States with quasienergy $E=0$ and $\pi/T$ appear in the spectrum with unit probability when $\nu_0$ and $\nu_{\pi}$ are nonzero, respectively, which distinguishes them from other states localized near the boundary. Hence the topological phases remain well defined in the presence of disorder.

We now analyze the case $E=0$. Repeatedly applying $T_i(0)$ to the unit cell amplitude $( \psi_{A1},\psi_{B1}) = (1,0)$ produces the unit cell amplitude $(\psi_{AN}, \psi_{BN}) =(\psi_{AN},0)$ whose magnitude decays at rate
\begin{equation}
\gamma_0 = \lim_{N \to \infty} \frac{1}{N} \log \Bigg|  \prod_{i=1}^N \frac{\sin   \frac{1}{2} {J}_iT}{\cos \frac{1}{2} {J}_{i+1}T} \cot \tfrac{1}{2} {K}_iT  \Bigg |,
\label{eq:lyap}
\end{equation}
where $\gamma_0$ is the signed Lyapunov exponent, or inverse localization length, of the state with energy quasienergy $0$. 

An edge state exists if $\gamma_0 < 0$ so that the wavefunction is normalizable, with the transition between trivial and topological phases occurring at $\gamma_0 = 0$. Since the localization length is the inverse of the absolute value of the Lyapunov exponent, we see that a change in $\nu_0$ is accompanied by a divergence in the localization length of states with quasienergy $0$. Likewise, a change in $\nu_{\pi}$ is accompanied by a divergence in the localization length of states with quasienergy $\pi/T$. We will discuss this divergence later on.

The product (\ref{eq:lyap}) can be split into a sum of three averages: an average of $\{\log|\sin \tfrac{1}{2} {J}_iT|\}$, $\{\log |\cos \tfrac{1}{2} {J}_{i+1}T| \}$, and $\{  \log|\cot \tfrac{1}{2} {K}_iT| \}$. Since each average is over independent, identically distributed random variables it converges to its corresponding statistical mean as we send $N \to \infty$. Letting an $\langle \cdots \rangle$ denote an expectation value over the disorder distribution, we find that 
\begin{subequations}
\label{eq:nu0}
\begin{align}
\label{eq:nu0_a}
\nu_0 &= 
\begin{cases}
1 & \text{if } \gamma_0   <0 \\
0 & \text{if } \gamma_0  >0
\end{cases} \\
\label{eq:nu0_b}
\gamma_0 &=   \langle\log|\tan \tfrac{1}{2}J_iT| \rangle - \langle\log|\tan \tfrac{1}{2}K_iT| \rangle.
\end{align}
\end{subequations}
An identical analysis of the case $E=\pi/T$ leads to 
\begin{subequations}
\label{eq:nupi}
\begin{align}
\nu_{\pi} &= 
\begin{cases}
-1 & \text{if } \gamma_{\pi}   <0 \\
0 & \text{if } \gamma_{\pi}  >0
\end{cases} \\
\gamma_{\pi} &=   - \langle\log|\tan \tfrac{1}{2}J_iT| \rangle - \langle\log|\tan \tfrac{1}{2}K_iT| \rangle.
\end{align}
\end{subequations}
As a check on these equations, consider the zero period limit, where $\mathcal{H}_{\text{eff}}$ reduces to the average of $\mathcal{H}(t)$ over a period \cite{Bukov15}. This average is exactly the Su-Schrieffer-Heeger model (\ref{eq:H_ti}), which supports zero modes whenever $\langle\log |{J_i}/{K_i} | \rangle <0$ \cite{Mondragon-Shem14}. Indeed, (\ref{eq:nu0}) reduces to this condition at small $T$.

\section{\label{phase_diagram}Phase diagram}

We now explore the phase diagram implied by (\ref{eq:nu0}) and (\ref{eq:nupi}). We will fix $T$ and vary the remaining energy scales. 

\begin{figure}[hbt]
\includegraphics{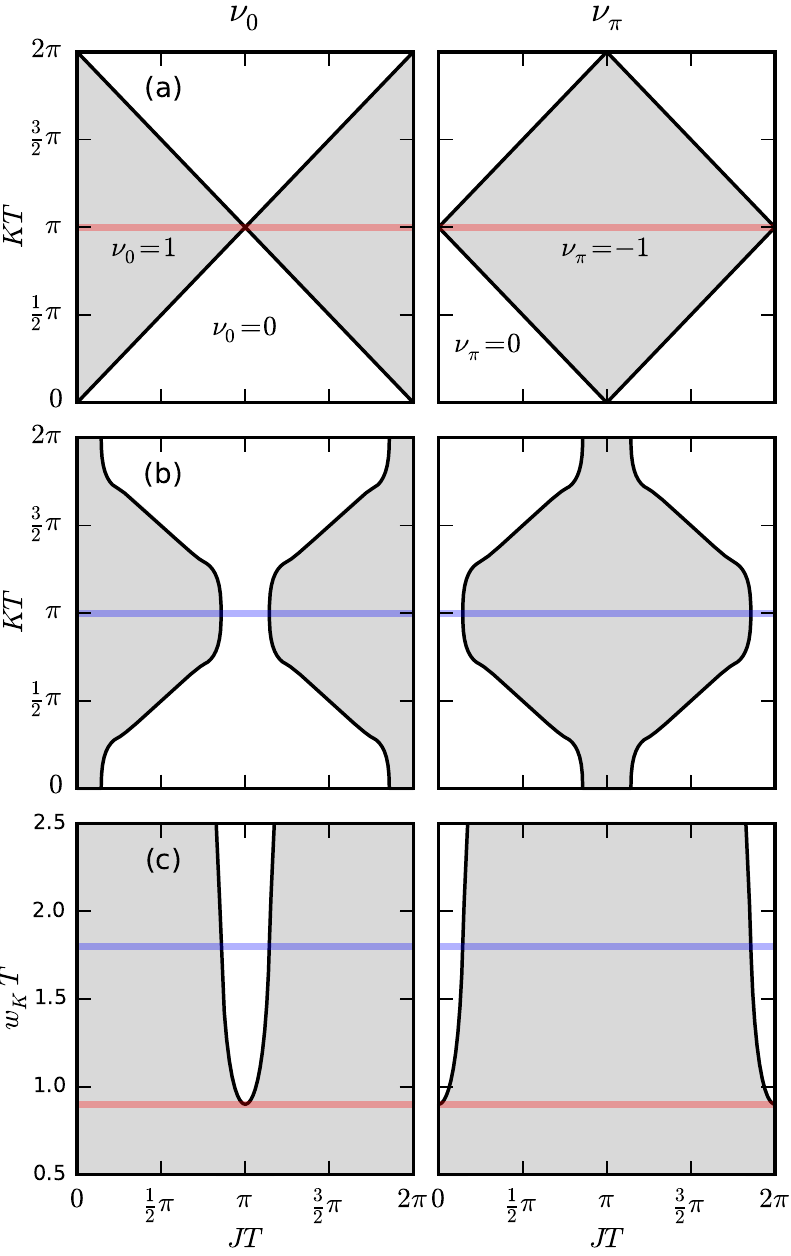}
\caption{\label{fig:analytic_phase} Phase diagram from analytic results. Left and right columns correspond to topological invariants $\nu_0$ and $\nu_{\pi}$, respectively. Panel (a): $(JT,KT)$ plane for zero disorder and for $wT = w_J T = w_K T < 2\pi$. Panel (b): $(JT,KT)$ plane for $w_J T = 0.9$, $w_K T = 1.8$.   Panel (c): $(JT, w_KT)$ plane for $KT=\pi$, $w_J T = 0.9$. The red and blue cuts from panels (a) and (b) respectively correspond to the red and blue cuts in panel (c), showing how the crossing of the phase boundaries present at $w_J T = w_K T = 0.9$ opens up on increasing $w_KT$.}
\end{figure}

In a clean system, $J_i$ and $K_i$ are distributed according to point distributions at $J$ and $K$, respectively. Then Equations (\ref{eq:nu0}) and (\ref{eq:nupi}) reduce to the known phase diagram \cite{Asboth14} for uniform bonds, shown in Fig~\ref{fig:analytic_phase}(a). The phase boundaries are located at
\begin{subequations}
\label{eq:clean_boundaries}
\begin{align}
JT \pm KT  &= 2n\pi \text{ for } \nu_0 \\
JT \pm KT &  = (2n+1)\pi \text{ for } \nu_{\pi}.
\end{align}
\end{subequations}

We now introduce disorder. For the sake of definiteness, we take $J_i = J+ w_J r_i$ and $K_i = K+ w_K q_i$, where $r_i$  and $q_i$ are random variables uniformly distributed in $[-\frac{1}{2},\frac{1}{2}]$. It is convenient to introduce the function
\begin{equation}
\label{eq:G_def}
G(x)  = \int_0^{x} du \log |\tan(u/2)|. 
\end{equation}
which has period $2\pi$ and is antisymmetric about $0$ and symmetric about $\pi/2$. In terms of $G$,
\begin{subequations}
\label{eq:lyap_full}
\begin{align}
\gamma_{0}&= \tfrac{1}{w_J T}\big[G(JT + \tfrac{1}{2}w_J T)  - G(JT - \tfrac{1}{2}w_J T)\big] \nonumber \\
&-\tfrac{1}{w_K T}\big[G(KT + \tfrac{1}{2}w_K T)  - G(KT - \tfrac{1}{2}w_K T)\big] \\
\gamma_{\pi}&= -\tfrac{1}{w_J T}\big[G(JT + \tfrac{1}{2}w_J T)  - G(JT - \tfrac{1}{2}w_J T)\big] \nonumber \\
&-\tfrac{1}{w_K T}\big[G(KT + \tfrac{1}{2}w_K T)  - G(KT - \tfrac{1}{2}w_K T)\big]. 
\end{align}
\end{subequations}

The simplest case to consider is $w_J=w_K=w$. Starting from the clean system, we increase $w$ and ask what happens to the phase boundaries in the $(JT,KT)$ plane. Using (\ref{eq:lyap_full}) and the fact that $G(x+2\pi) =  G(x)$ and $G(x+ \pi) = -G(x)$, we see that $\gamma_0$ and $\gamma_{\pi}$ continue to vanish along the lines defined by (\ref{eq:clean_boundaries}) even for nonzero $w$. Hence the phase boundaries remain fixed up until $wT = 2\pi$, at which point $G(x+2\pi) = G(x)$ implies that the entire $(JT,KT)$ plane becomes critical and $\gamma_0$ and $\gamma_{\pi}$ switch signs. This signals a disorder induced phase transition at which every point in the $(JT,KT)$ plane simultaneously switches from trivial to topological, or vice versa. 

There are infinitely many such disorder induced transitions, corresponding to $wT= 2n\pi$ for integer $n$. These occur because the time evolution operator (\ref{eq:explicit_UT}) depends only on $\widetilde{J_iT} = J_iT \mod 2\pi$ and $\widetilde{K_iT} = K_iT \mod 2\pi$, and whenever $wT= 2n\pi$ the distribution for $\widetilde{J_iT}$ and $\widetilde{K_iT}$ returns to itself.

Next we can ask what happens if $w_J \ne w_K$. Consider $w_JT, w_KT < 2\pi$ and $w_K > w_J$. In this case the phase boundaries shift in the $(JT,KT)$ plane relative to their $w=0$ positions. The qualitative distinction at moderate disorder is that the crossing of the  $\nu_0$ phase boundaries at $(JT,KT) = (0,0), (\pi,\pi)$ and the $\nu_{\pi}$ phase boundaries at $(JT,KT) = (0,\pi),(\pi,0)$ open up. For instance, at $(JT,KT)=(\pi,\pi)$
\begin{align}
\gamma_0 &=  - \frac{2}{w_JT} G(\tfrac{1}{2}w_JT) + \frac{2}{w_KT}G(\tfrac{1}{2}w_KT) \nonumber \\
& > 0 \text{ when } w_K > w_V
\end{align}
so the crossing opens horizontally. An explicit example is shown in Figures \ref{fig:analytic_phase}(b) and (c). If we continue to increase $w_KT$ up to $2\pi$, the phase boundaries flatten out into vertical lines. If we instead consider $w_K <w_J$, then the crossings open vertically rather than horizontally.

\section{\label{rs_wind}Real-space Winding Number}

To supplement these analytic results, we also numerically compute $\nu_0$ and $\nu_{\pi}$ in terms of a real-space winding number. 

The general expression for $\nu_0$ and $\nu_{\pi}$ in Floquet systems exhibiting chiral symmetry was worked out by Refs.~\onlinecite{Asboth13, Asboth14}: let $\mathcal{H}_{\text{eff}}$ be the effective Hamiltonian of the system, and let $\mathcal{H}_{\text{eff}}'$ be the effective Hamiltonian for a version of the drive shifted in starting time by $\delta T \ne nT$. The shift $\delta T$ is chosen so that $\mathcal{H}_{\text{eff}}'$ also respects chiral symmetry; for our system $\delta T = \frac{1}{2}T$\cite{Asboth13,Asboth14}. Then
\begin{subequations}
\begin{align}
\nu_0 &= \frac{\nu + \nu'}{2} \\
\nu_{\pi} &= \frac{\nu - \nu'}{2},
\end{align}
\end{subequations}
Here $\nu \in \mathbb{Z}$ is the topological invariant for the time-independent matrix $\mathcal{H}_{\text{eff}}$ and $\nu'$ the corresponding invariant for $\mathcal{H}_{\text{eff}}'$. 

\begin{figure}[t]
\includegraphics{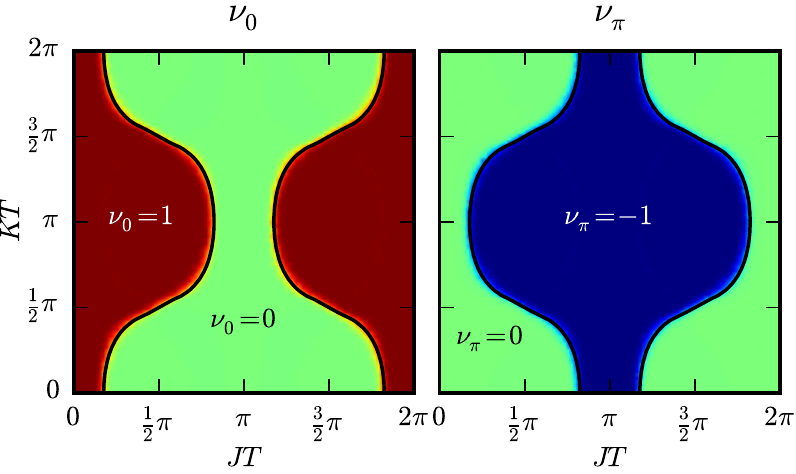}
\caption{\label{fig:winding_all} Phase diagram from numeric computation of the real space winding number. Plots of $\nu_0$ (left) and $\nu_{\pi}$ (right), in the $(JT,KT)$ plane for fixed disorder $w_J T = 0.5$, $w_K T = 2.8$. The black lines are the analytically obtained phase boundaries. The computation used $101$ unit cells and $100$ disorder realizations. }
\end{figure}

For time-independent systems belonging to class BDI, $\nu \in \mathbb{Z}$ is given by the following winding number whenever translational symmetry is available:
\begin{equation}
\label{eq:winding}
\nu = \frac{-1}{2\pi i} \int dk  \hspace{1.5 pt} \text{tr} \Big( h^{-1}(k) \frac{d}{dk} h(k) \Big ) 
\end{equation}
Here $h$ is the off diagonal block of the Bloch Hamiltonian $\mathcal{H}(k)$, as obtained when $\mathcal{H}(k)$ is written in the basis where $\mathcal{S}$ is diagonal \cite{Ryu10, Song14, Mondragon-Shem14, Asboth14}. Refs.~\onlinecite{Mondragon-Shem14,Song14} have generalized the notion of a winding number to disordered systems by expressing (\ref{eq:winding}) in real space. Following the procedure developed in their work, we numerically compute $\nu$ and $\nu'$, thus obtaining $\nu_0$ and $\nu_{\pi}$. The results are shown in Fig. \ref{fig:winding_all} and are in full agreement with the analytic results derived above.

\section{\label{crit_exp}Localization Lengths and Critical Exponents}

One dimensional systems are expected to localize in the presence of disorder \cite{Kramer93}. To check that this holds for our Floquet system, we have used the full transfer matrix (\ref{eq:T}) to numerically compute the localization length of the system as a function of quasienergy. Except for states with quasienergies $0$ and $\pi/T$ which delocalize when $\nu_0$ and $\nu_{\pi}$ change, respectively, all states are indeed localized in the presence of disorder. Therefore the system is a promising candidate for being many-body localized in the presence of interactions.

We now examine the transition between distinct localized phases, deriving the critical exponents governing the divergence of the localization length of states with quasienergy $0$ and $\pi/T$ at phase transitions. We do this by finding the form of (\ref{eq:lyap_full}) near a phase transition. We present the result here, with a careful analysis in appendix \ref{app:loc_length}.

Let $g$ represent one of $JT$, $KT$, $w_JT$, or $ w_KT$ and $g_c$ the corresponding critical value. Let $l_0$ and $l_{\pi}$ denote the localization lengths of states with quasienergy $0$ and $\pi/T$, respectively. Then generically
\begin{subequations}
\label{eq:crit_exp}
\begin{align}
l_{0} & \sim  |g-g_c|^{-1} \text{ when $\nu_0$ changes} \\
l_{\pi} & \sim  |g-g_c|^{-1} \text{ when $\nu_{\pi}$ changes}.
\end{align}
\end{subequations}
Anomalous scaling of the form
\begin{subequations}
\label{eq:crit_exp1}
\begin{align}
l_{0} & \sim  \big| (g-g_c) \log{|g-g_c|} \big|^{-1} \text{ when $\nu_0$ changes} \\
l_{\pi} & \sim  \big| (g-g_c) \log{|g-g_c|} \big|^{-1} \text{ when $\nu_{\pi}$ changes}
\end{align}
\end{subequations}
appears upon tuning $JT$ or $w_JT$ when $(JT)_c = \pm \frac{1}{2} (w_JT)_c +n\pi$, and upon tuning  $KT$ or $w_KT$ when $(KT)_c = \pm \frac{1}{2} (w_KT)_c +n\pi$. The region of anomalous scaling has codimension $2$ and is therefore not a generic transition point. Both the standard and anomalous scaling match results for the time independent system $(\ref{eq:H_ti})$ \cite{Mondragon-Shem14}. This is notable because the phase transition where $\nu_{\pi}$ changes is unique to Floquet systems.

\section{\label{conclusion}Discussion}
To summarize, we have found that disorder shifts the topological phase boundaries relative to the clean system according to Equations (\ref{eq:nu0}) and (\ref{eq:nupi}). It is also noteworthy that the domain of application of the real-space winding number has been extended to include Floquet systems. In addition we found that the exponent for the divergence of the localization length of topological edge states at phase transitions takes on equilibrium values, despite the fact that phases with nonzero $\nu_{\pi}$ are unique to equilibrium. We confirmed that the system is localized away from these transitions. Hence, the overall picture is very similar to that for static systems belonging to class BDI.

We also mention that our work is closely connected to recent work by V. Khemani, A. Lazarides, R. Moessner, and S. L. Sondhi \cite{Khemani15}, in which the authors consider a driven Ising model and show that disorder protects Floquet spin-glass order against interactions. They discuss how the model that they consider can be mapped to a Floquet superconductor that exhibits Majorana edge modes with energy $0$ and $\pi/T$. In the noninteracting limit, the BdG Hamiltonian describing this superconductor system maps onto the Hamiltonian that we consider in this work, up to an intercell rotation by $\frac{1}{\sqrt{2}}(1+ \sigma_z) \otimes 1$. They add weak interactions to a point corresponding to $(JT,KT) = (3.06,1.885)$, $(w_JT,w_KT) = (0.078,2.198)$ in our phase diagram, finding that the so called $\pi$ spin-glass corresponding to $\pi/T$ Majoranas survives.

The natural question for future work is whether the analogous result holds for our formulation of the problem; that is, whether the many-body system described by (\ref{eq:many_body}) supports edge modes when weak interactions are included.

We would like to acknowledge that while finalizing this manuscript we became aware of a previous work by T. Rakovszky and J. K. Asboth \cite{Rakovszky15}, in which the authors derive equivalent results for the quantum walk analogue of the Floquet system studied here.

\section{\label{Acknowledgements}Acknowledgements}
I would like to thank A. Vishwanath for introducing me to this problem and for helpful discussions and comments. I would also like to thank P.T. Dumitrescu, H. C. Po, A. C. Potter, and N. Yao for helpful discussions and comments.

\appendix

\section{\label{app:computation}Details of Transfer Matrix Computation}
In this Appendix we outline the steps leading from the Floquet eigenvalue equation $\mathcal{U}_T\psi = e^{-iET}\psi$ to the transfer matrix (\ref{eq:T_form}). It is useful to instead consider the operator $\mathcal{U}_{T}' = e^{-\frac{i}{2}T\mathcal{H}_2(\{K_i\})}e^{-\frac{i}{2}T\mathcal{H}_1(\{J_i\})}$ which satisfies $\mathcal{U}_T ' \psi ' = e^{-iET}\psi ' $, where $\psi ' = \Lambda^{\dagger} \psi $ and $\Lambda= e^{-\frac{i}{4}T\mathcal{H}_1(\{J_i\})}$. This transformation minimizes the range of the matrix involved. Physically, it is equivalent to a shift in starting time of the periodic drive.

Letting site index $Ai$ correspond to matrix index $2i-1$ and site index $Bi$ to matrix index $2i$,

\begin{widetext}
\begin{equation}
\label{eq:floq_matrix}
\begin{pmatrix*}[c]
F_1		&	f_1		&	0		&	0		&	0		& 	0		&	 \cdots\\
f_1G_1	& 	F_1G_1 	& 	F_2g_1 	&	f_2g_1 	&	0  		&	0		&	\cdots\\
f_1g_1	& 	F_1g_1 	&	F_2G_1 	&	f_2G_1 	&	0  		&	0		&	 \cdots\\
0 		& 	0		& 	f_2G_2	& 	F_2G_2 	&	F_3g_2  	&	f_3g_2	&	 \cdots\\
0 		&	0		& 	f_2g_2	& 	F_2g_2	& 	F_3G_2 	&	f_3G_2	&	 \cdots\\	
\vdots	&	\vdots	&	\vdots	&	\vdots	&	\vdots	&	\vdots	& 	 \ddots\\
\end{pmatrix*}
\begin{pmatrix}
\psi_{A1}' \\
\psi_{B1}' \\
\psi_{A2}' \\
\psi_{B2}' \\
\psi_{A3}' \\
\vdots
\end{pmatrix}
=
e^{-iET}
\begin{pmatrix}
\psi_{A1}' \\
\psi_{B1}' \\
\psi_{A2}' \\
\psi_{B2}' \\
\psi_{A3}' \\
\vdots
\end{pmatrix}
\end{equation}
\end{widetext}
where $F_i = \cos{J_iT}$,$f_i = -i\sin{J_iT}$, $G_i =\cos{K_iT}$, and $g_i = -i\sin{K_iT}$. Examining rows $2i$ and $2i+1$ of (\ref{eq:floq_matrix}), we find
\begin{equation}
M_i 
\begin{pmatrix}
\psi_{Ai}' \\
\psi_{Bi}'
\end{pmatrix}
= N_i
\begin{pmatrix}
\psi_{Ai+1}' \\
\psi_{Bi+1}'
\end{pmatrix}
\end{equation}
where
\begin{subequations}
\begin{align}
M_i =  &
\begin{pmatrix}
f_i G_i	&	F_i G_i - e^{-iET} \\
f_i g_i	&	F_i g_i
\end{pmatrix} \\
N_i = &
\begin{pmatrix}
-F_{i+1} g_i	&	-f_{i+1} g_i \\
e^{-iET}-F_{i+1}G_i		&	-f_{i+1}G_i \\
\end{pmatrix}.
\end{align}
\end{subequations}
In order to switch back to a transfer matrix equation for $\mathcal{U}_T$, note that $\Lambda$ is block diagonal with blocks separated by unit cells. Letting $\Lambda_i$ denote the $i$th block of this matrix, 
\begin{equation}
\Lambda_i =
\begin{pmatrix}
\cos{\frac{1}{2}J_iT} & -i\sin{\frac{1}{2}J_iT} \\
 -i\sin{\frac{1}{2}J_iT} & \cos{\frac{1}{2}J_iT} 
\end{pmatrix},
\end{equation}
this implies
\begin{equation}
M_i  \Lambda_{i}^{\dagger}
\begin{pmatrix}
\psi_{Ai}\\
\psi_{Bi}
\end{pmatrix}
= N_i \Lambda_{i+1}^{\dagger}
\begin{pmatrix}
\psi_{Ai+1} \\
\psi_{Bi+1}
\end{pmatrix},
\end{equation}
so that 
\begin{equation}
T_i(E) = \Lambda_{i+1} N_i^{-1} M_i \Lambda_{i}^{\dagger}.
\end{equation}
Plugging in $F_i$, $G_i$, $f_i$, $g_i$ and performing the matrix multiplication we find
\begin{widetext}
\begin{equation}
\label{eq:T}
T_i(E) = \frac{1}{ \sin K_iT}
\begin{pmatrix*}[c]
-\dfrac{\sin   \frac{1}{2} J_iT}{\cos \frac{1}{2} J_{i+1}T} (\cos ET +  \cos K_iT) &  \dfrac{\cos \frac{1}{2} J_iT}{\cos \frac{1}{2} J_{i+1}T} \sin ET \\
- \dfrac{\sin \frac{1}{2} J_iT}{\sin  \frac{1}{2} J_{i+1}T} \sin ET & -\dfrac{\cos \frac{1}{2} J_iT}{\sin  \frac{1}{2} J_{i+1}T} (\cos ET  -  \cos K_iT)
\end{pmatrix*} .
\end{equation}
\end{widetext}

The boundary conditions quoted in section \ref{transfer_matrix} are derived by combining the very first row of (\ref{eq:floq_matrix}) for $E=0, \pi/T$ with the equations for $\psi_{A1'}$ and $\psi_{B1'}$ in terms of $\psi_{A1}$ and $\psi_{B1}$.

\section{\label{app:loc_length}Localization Length}
In this appendix we derive Equations (\ref{eq:crit_exp}) and (\ref{eq:crit_exp1}) for the divergence of localization lengths, using an analysis similar to Ref. \onlinecite{Mondragon-Shem14}. For definiteness, we examine the localization length $l_0$ of states with quasienergy $0$, and we consider tuning one of $JT$, $w_J T$ with the remaining parameters fixed. The analysis is the same if we consider $l_{\pi}$ instead, or if we tune one of $KT$, $w_KT$.

It is useful to introduce an additional function
\begin{equation}
\label{eq:F}
F(x,a) = G(x+a) - G(x-a)
\end{equation}
where $G(x)$ is defined in (\ref{eq:G_def}). Setting $(x,a) = (JT, \frac{1}{2}w_J T)$ and using (\ref{eq:lyap_full}),
\begin{equation}
\label{lyap_F}
\gamma_0 = \pm \frac{1}{2a} F(x, a) - \frac{1}{w_KT} F(KT, \tfrac{1}{2}w_KT).
\end{equation}
Let $(x_c, a_c) = \big((JT)_c, \frac{1}{2}(w_JT)_c \big)$ correspond to a critical point where $\nu_0$ changes value, and let $(\delta x,\delta a) = (x- x_c, a-a_c)$. At the transition, $\gamma_0=0$, so $\gamma_0$ is given by the non-constant piece of the expansion of  $\frac{1}{2a} F(x, a)$ in terms of $\delta x$ and $\delta a$. 

Combining (\ref{eq:G_def}) and (\ref{eq:F}), 
\begin{subequations}
\begin{align}
\frac{\partial}{\partial x} F(x,a_c) &= \log |\sin{x} + \sin{a_c}| - |\sin{x} - \sin{a_c}| \\
\frac{\partial}{\partial a} F(x_c,a) &= \log |\cos{a} - \cos{x_c}| - |\cos{a} + \cos{x_c}| .
\end{align}
\end{subequations}
As long as $x_c \ne \pm a_c + n\pi$, both functions above are analytic at $(x_c, a_c)$ and can be expanded in a taylor series. Integrating then gives
$\frac{1}{a_c} F(x,a_c)$ in terms of $\delta x$ 
\begin{equation}
\label{eq:tune_x}
\frac{1}{a_c} F(x,a_c) = \frac{1}{a_c} \bigg[ F(x_c,a_c) + C_1 \delta x + \tfrac{1}{2}C_2 \delta x^2 + \mathcal{O} (\delta x ^3) \bigg],
\end{equation}
and $\frac{1}{a} F(x_c,a)$ in terms of $\delta a$
\begin{align}
\label{eq:tune_a}
\frac{1}{a} F(x_c,a) = \frac{1}{a_c} \bigg[ F(x_c,a_c) + \Big(D_1 -\tfrac{F(x_c,a_c)}{a_c} \Big) \delta a \nonumber \\
 + \Big( \tfrac{1}{2}D_2 + \tfrac{F(x_c,a_c)}{a_c^2} - \tfrac{D_1}{a_c} \Big)  \delta a^2 + \mathcal{O} (\delta a ^3) \bigg],
\end{align}
where 
\begin{subequations}
\begin{align}
C_1 &= \log |\sin{x_c} + \sin{a_c}| - |\sin{x_c} - \sin{a_c}| \\
D_1 &=  \log |\cos{a_c} - \cos{x_c}| - |\cos{a_c} + \cos{x_c}| \\
C_2 &= 2 \frac{\cos{x_c} \sin{a_c}}{\sin^2{a_c} - \sin^2{x_c}} \\
D_2 &=  2 \frac{\cos{x_c} \sin{a_c}}{\cos^2{x_c} - \cos^2{a_c}}.
\end{align}
\end{subequations}

If the linear coefficient in (\ref{eq:tune_x}) vanishes, then either the quadratic coefficient is nonzero or $F(x,a_c)$ is constant. Both scenarios correspond to a trajectory in parameter space that is tangent to the phase boundaries, which is not a phase transition. The same holds for (\ref{eq:tune_a}). Hence, at a true transition
\begin{subequations}
\begin{align}
\gamma_0 &\sim JT - (JT)_c \text{ on tuning $JT$} \\
\gamma_0 &\sim w_JT - (w_JT)_c \text{ on tuning $w_JT$}.
\end{align}
\end{subequations}
Since $l_0 = |\gamma_0|^{-1}$, this gives
\begin{subequations}
\begin{align}
\l_0 &\sim \big|JT - (JT)_c  \big|^{-1} \text{ on tuning $JT$} \\
\l_0 &\sim \big| w_JT - (w_JT)_c \big|^{-1} \text{ on tuning $w_JT$}.
\end{align}
\end{subequations}
Now consider $x_c = \pm a_c + n\pi$, which corresponds to $(JT)_c = \pm \frac{1}{2} (w_JT)_c + n\pi$. Then 
\begin{subequations}
\begin{align}
\frac{\partial}{\partial x} F(x,a_c) &= \tilde{C} \log|\delta x| + \text{powers of $\delta x$}  \\
\frac{\partial}{\partial a} F(x_c,a) &= \tilde{D} \log|\delta a| + \text{powers of $\delta a$}
\end{align}
\end{subequations}
for nonzero $\tilde{C}$ and $\tilde{D}$. Integrating this we find that $\frac{1}{a_c} F(x,a_c)$ and $\frac{1}{a} F(x_c,a)$ have leading non-constant terms proportional to $ \delta x\log|\delta x|$ and $ \delta a\log |\delta a|$, respectively. Thus
\begin{subequations}
\begin{align}
\l_0 &\sim \big| \big( JT - (JT)_c \big) \log| JT - (JT)_c |  \big|^{-1} \nonumber \\ & \text{ on tuning $JT$} \\
\l_0 &\sim \big| \big(w_JT - (w_JT)_c \big) \log |  w_JT - (w_JT)_c |   \big|^{-1} \nonumber \\ & \text{ on tuning $w_JT$}.
\end{align}
\end{subequations}
This is the anomalous scaling in (\ref{eq:crit_exp1}).


\bibliography{bibliography}

\providecommand{\noopsort}[1]{}\providecommand{\singleletter}[1]{#1}%
\begin{thebibliography}{39}%
\makeatletter
\providecommand \@ifxundefined [1]{%
 \@ifx{#1\undefined}
}%
\providecommand \@ifnum [1]{%
 \ifnum #1\expandafter \@firstoftwo
 \else \expandafter \@secondoftwo
 \fi
}%
\providecommand \@ifx [1]{%
 \ifx #1\expandafter \@firstoftwo
 \else \expandafter \@secondoftwo
 \fi
}%
\providecommand \natexlab [1]{#1}%
\providecommand \enquote  [1]{``#1''}%
\providecommand \bibnamefont  [1]{#1}%
\providecommand \bibfnamefont [1]{#1}%
\providecommand \citenamefont [1]{#1}%
\providecommand \href@noop [0]{\@secondoftwo}%
\providecommand \href [0]{\begingroup \@sanitize@url \@href}%
\providecommand \@href[1]{\@@startlink{#1}\@@href}%
\providecommand \@@href[1]{\endgroup#1\@@endlink}%
\providecommand \@sanitize@url [0]{\catcode `\\12\catcode `\$12\catcode
  `\&12\catcode `\#12\catcode `\^12\catcode `\_12\catcode `\%12\relax}%
\providecommand \@@startlink[1]{}%
\providecommand \@@endlink[0]{}%
\providecommand \url  [0]{\begingroup\@sanitize@url \@url }%
\providecommand \@url [1]{\endgroup\@href {#1}{\urlprefix }}%
\providecommand \urlprefix  [0]{URL }%
\providecommand \Eprint [0]{\href }%
\providecommand \doibase [0]{http://dx.doi.org/}%
\providecommand \selectlanguage [0]{\@gobble}%
\providecommand \bibinfo  [0]{\@secondoftwo}%
\providecommand \bibfield  [0]{\@secondoftwo}%
\providecommand \translation [1]{[#1]}%
\providecommand \BibitemOpen [0]{}%
\providecommand \bibitemStop [0]{}%
\providecommand \bibitemNoStop [0]{.\EOS\space}%
\providecommand \EOS [0]{\spacefactor3000\relax}%
\providecommand \BibitemShut  [1]{\csname bibitem#1\endcsname}%
\let\auto@bib@innerbib\@empty
\bibitem [{\citenamefont {Hasan}\ and\ \citenamefont {Kane}(2010)}]{Hasan10}%
  \BibitemOpen
  \bibfield  {author} {\bibinfo {author} {\bibfnamefont {M.~Z.}\ \bibnamefont
  {Hasan}}\ and\ \bibinfo {author} {\bibfnamefont {C.~L.}\ \bibnamefont
  {Kane}},\ }\href {\doibase 10.1103/RevModPhys.82.3045} {\bibfield  {journal}
  {\bibinfo  {journal} {Rev. Mod. Phys.}\ }\textbf {\bibinfo {volume} {82}},\
  \bibinfo {pages} {3045} (\bibinfo {year} {2010})}\BibitemShut {NoStop}%
\bibitem [{\citenamefont {Qi}\ and\ \citenamefont {Zhang}(2011)}]{Qi10}%
  \BibitemOpen
  \bibfield  {author} {\bibinfo {author} {\bibfnamefont {X.-L.}\ \bibnamefont
  {Qi}}\ and\ \bibinfo {author} {\bibfnamefont {S.-C.}\ \bibnamefont {Zhang}},\
  }\href {\doibase 10.1103/RevModPhys.83.1057} {\bibfield  {journal} {\bibinfo
  {journal} {Rev. Mod. Phys.}\ }\textbf {\bibinfo {volume} {83}},\ \bibinfo
  {pages} {1057} (\bibinfo {year} {2011})}\BibitemShut {NoStop}%
\bibitem [{\citenamefont {Senthil}(2015)}]{Senthil15}%
  \BibitemOpen
  \bibfield  {author} {\bibinfo {author} {\bibfnamefont {T.}~\bibnamefont
  {Senthil}},\ }\href {\doibase 10.1146/annurev-conmatphys-031214-014740}
  {\bibfield  {journal} {\bibinfo  {journal} {Annual Review of Condensed Matter
  Physics}\ }\textbf {\bibinfo {volume} {6}},\ \bibinfo {pages} {299} (\bibinfo
  {year} {2015})},\ \Eprint
  {http://arxiv.org/abs/http://dx.doi.org/10.1146/annurev-conmatphys-031214-014740}
  {http://dx.doi.org/10.1146/annurev-conmatphys-031214-014740} \BibitemShut
  {NoStop}%
\bibitem [{\citenamefont {Oka}\ and\ \citenamefont {Aoki}(2009)}]{Takashi09}%
  \BibitemOpen
  \bibfield  {author} {\bibinfo {author} {\bibfnamefont {T.}~\bibnamefont
  {Oka}}\ and\ \bibinfo {author} {\bibfnamefont {H.}~\bibnamefont {Aoki}},\
  }\href {\doibase 10.1103/PhysRevB.79.081406} {\bibfield  {journal} {\bibinfo
  {journal} {Phys. Rev. B}\ }\textbf {\bibinfo {volume} {79}},\ \bibinfo
  {pages} {081406} (\bibinfo {year} {2009})}\BibitemShut {NoStop}%
\bibitem [{\citenamefont {Kitagawa}\ \emph
  {et~al.}(2010{\natexlab{a}})\citenamefont {Kitagawa}, \citenamefont {Berg},
  \citenamefont {Rudner},\ and\ \citenamefont {Demler}}]{Kitagawa10}%
  \BibitemOpen
  \bibfield  {author} {\bibinfo {author} {\bibfnamefont {T.}~\bibnamefont
  {Kitagawa}}, \bibinfo {author} {\bibfnamefont {E.}~\bibnamefont {Berg}},
  \bibinfo {author} {\bibfnamefont {M.}~\bibnamefont {Rudner}}, \ and\ \bibinfo
  {author} {\bibfnamefont {E.}~\bibnamefont {Demler}},\ }\href {\doibase
  10.1103/PhysRevB.82.235114} {\bibfield  {journal} {\bibinfo  {journal} {Phys.
  Rev. B}\ }\textbf {\bibinfo {volume} {82}},\ \bibinfo {pages} {235114}
  (\bibinfo {year} {2010}{\natexlab{a}})}\BibitemShut {NoStop}%
\bibitem [{\citenamefont {Jiang}\ \emph {et~al.}(2011)\citenamefont {Jiang},
  \citenamefont {Kitagawa}, \citenamefont {Alicea}, \citenamefont {Akhmerov},
  \citenamefont {Pekker}, \citenamefont {Refael}, \citenamefont {Cirac},
  \citenamefont {Demler}, \citenamefont {Lukin},\ and\ \citenamefont
  {Zoller}}]{Jiang11}%
  \BibitemOpen
  \bibfield  {author} {\bibinfo {author} {\bibfnamefont {L.}~\bibnamefont
  {Jiang}}, \bibinfo {author} {\bibfnamefont {T.}~\bibnamefont {Kitagawa}},
  \bibinfo {author} {\bibfnamefont {J.}~\bibnamefont {Alicea}}, \bibinfo
  {author} {\bibfnamefont {A.~R.}\ \bibnamefont {Akhmerov}}, \bibinfo {author}
  {\bibfnamefont {D.}~\bibnamefont {Pekker}}, \bibinfo {author} {\bibfnamefont
  {G.}~\bibnamefont {Refael}}, \bibinfo {author} {\bibfnamefont {J.~I.}\
  \bibnamefont {Cirac}}, \bibinfo {author} {\bibfnamefont {E.}~\bibnamefont
  {Demler}}, \bibinfo {author} {\bibfnamefont {M.~D.}\ \bibnamefont {Lukin}}, \
  and\ \bibinfo {author} {\bibfnamefont {P.}~\bibnamefont {Zoller}},\ }\href
  {\doibase 10.1103/PhysRevLett.106.220402} {\bibfield  {journal} {\bibinfo
  {journal} {Phys. Rev. Lett.}\ }\textbf {\bibinfo {volume} {106}},\ \bibinfo
  {pages} {220402} (\bibinfo {year} {2011})}\BibitemShut {NoStop}%
\bibitem [{\citenamefont {Kitagawa}\ \emph {et~al.}(2011)\citenamefont
  {Kitagawa}, \citenamefont {Oka}, \citenamefont {Brataas}, \citenamefont
  {Fu},\ and\ \citenamefont {Demler}}]{Kitagawa11b}%
  \BibitemOpen
  \bibfield  {author} {\bibinfo {author} {\bibfnamefont {T.}~\bibnamefont
  {Kitagawa}}, \bibinfo {author} {\bibfnamefont {T.}~\bibnamefont {Oka}},
  \bibinfo {author} {\bibfnamefont {A.}~\bibnamefont {Brataas}}, \bibinfo
  {author} {\bibfnamefont {L.}~\bibnamefont {Fu}}, \ and\ \bibinfo {author}
  {\bibfnamefont {E.}~\bibnamefont {Demler}},\ }\href {\doibase
  10.1103/PhysRevB.84.235108} {\bibfield  {journal} {\bibinfo  {journal} {Phys.
  Rev. B}\ }\textbf {\bibinfo {volume} {84}},\ \bibinfo {pages} {235108}
  (\bibinfo {year} {2011})}\BibitemShut {NoStop}%
\bibitem [{\citenamefont {Lindner}\ \emph {et~al.}(2011)\citenamefont
  {Lindner}, \citenamefont {Refael},\ and\ \citenamefont
  {Galitski}}]{Lindner11}%
  \BibitemOpen
  \bibfield  {author} {\bibinfo {author} {\bibfnamefont {N.~H.}\ \bibnamefont
  {Lindner}}, \bibinfo {author} {\bibfnamefont {G.}~\bibnamefont {Refael}}, \
  and\ \bibinfo {author} {\bibfnamefont {V.}~\bibnamefont {Galitski}},\ }\href
  {http://dx.doi.org/10.1038/nphys1926} {\bibfield  {journal} {\bibinfo
  {journal} {Nat Phys}\ }\textbf {\bibinfo {volume} {7}},\ \bibinfo {pages}
  {490} (\bibinfo {year} {2011})}\BibitemShut {NoStop}%
\bibitem [{\citenamefont {Rudner}\ \emph {et~al.}(2013)\citenamefont {Rudner},
  \citenamefont {Lindner}, \citenamefont {Berg},\ and\ \citenamefont
  {Levin}}]{Rudner13}%
  \BibitemOpen
  \bibfield  {author} {\bibinfo {author} {\bibfnamefont {M.~S.}\ \bibnamefont
  {Rudner}}, \bibinfo {author} {\bibfnamefont {N.~H.}\ \bibnamefont {Lindner}},
  \bibinfo {author} {\bibfnamefont {E.}~\bibnamefont {Berg}}, \ and\ \bibinfo
  {author} {\bibfnamefont {M.}~\bibnamefont {Levin}},\ }\href {\doibase
  10.1103/PhysRevX.3.031005} {\bibfield  {journal} {\bibinfo  {journal} {Phys.
  Rev. X}\ }\textbf {\bibinfo {volume} {3}},\ \bibinfo {pages} {031005}
  (\bibinfo {year} {2013})}\BibitemShut {NoStop}%
\bibitem [{\citenamefont {Asb\'oth}\ \emph {et~al.}(2014)\citenamefont
  {Asb\'oth}, \citenamefont {Tarasinski},\ and\ \citenamefont
  {Delplace}}]{Asboth14}%
  \BibitemOpen
  \bibfield  {author} {\bibinfo {author} {\bibfnamefont {J.~K.}\ \bibnamefont
  {Asb\'oth}}, \bibinfo {author} {\bibfnamefont {B.}~\bibnamefont
  {Tarasinski}}, \ and\ \bibinfo {author} {\bibfnamefont {P.}~\bibnamefont
  {Delplace}},\ }\href {\doibase 10.1103/PhysRevB.90.125143} {\bibfield
  {journal} {\bibinfo  {journal} {Phys. Rev. B}\ }\textbf {\bibinfo {volume}
  {90}},\ \bibinfo {pages} {125143} (\bibinfo {year} {2014})}\BibitemShut
  {NoStop}%
\bibitem [{\citenamefont {{Nathan}}\ and\ \citenamefont
  {{Rudner}}(2015)}]{Nathan15}%
  \BibitemOpen
  \bibfield  {author} {\bibinfo {author} {\bibfnamefont {F.}~\bibnamefont
  {{Nathan}}}\ and\ \bibinfo {author} {\bibfnamefont {M.~S.}\ \bibnamefont
  {{Rudner}}},\ }\href@noop {} {\bibfield  {journal} {\bibinfo  {journal}
  {ArXiv e-prints}\ } (\bibinfo {year} {2015})},\ \Eprint
  {http://arxiv.org/abs/1506.07647} {arXiv:1506.07647 [cond-mat.mes-hall]}
  \BibitemShut {NoStop}%
\bibitem [{\citenamefont {Kitagawa}\ \emph
  {et~al.}(2010{\natexlab{b}})\citenamefont {Kitagawa}, \citenamefont {Rudner},
  \citenamefont {Berg},\ and\ \citenamefont {Demler}}]{Kitagawa10b}%
  \BibitemOpen
  \bibfield  {author} {\bibinfo {author} {\bibfnamefont {T.}~\bibnamefont
  {Kitagawa}}, \bibinfo {author} {\bibfnamefont {M.~S.}\ \bibnamefont
  {Rudner}}, \bibinfo {author} {\bibfnamefont {E.}~\bibnamefont {Berg}}, \ and\
  \bibinfo {author} {\bibfnamefont {E.}~\bibnamefont {Demler}},\ }\href
  {\doibase 10.1103/PhysRevA.82.033429} {\bibfield  {journal} {\bibinfo
  {journal} {Phys. Rev. A}\ }\textbf {\bibinfo {volume} {82}},\ \bibinfo
  {pages} {033429} (\bibinfo {year} {2010}{\natexlab{b}})}\BibitemShut
  {NoStop}%
\bibitem [{\citenamefont {{Kitagawa}}(2011)}]{Kitagawa11}%
  \BibitemOpen
  \bibfield  {author} {\bibinfo {author} {\bibfnamefont {T.}~\bibnamefont
  {{Kitagawa}}},\ }\href@noop {} {\bibfield  {journal} {\bibinfo  {journal}
  {ArXiv e-prints}\ } (\bibinfo {year} {2011})},\ \Eprint
  {http://arxiv.org/abs/1112.1882} {arXiv:1112.1882 [quant-ph]} \BibitemShut
  {NoStop}%
\bibitem [{\citenamefont {Asb\'oth}\ and\ \citenamefont
  {Obuse}(2013)}]{Asboth13}%
  \BibitemOpen
  \bibfield  {author} {\bibinfo {author} {\bibfnamefont {J.~K.}\ \bibnamefont
  {Asb\'oth}}\ and\ \bibinfo {author} {\bibfnamefont {H.}~\bibnamefont
  {Obuse}},\ }\href {\doibase 10.1103/PhysRevB.88.121406} {\bibfield  {journal}
  {\bibinfo  {journal} {Phys. Rev. B}\ }\textbf {\bibinfo {volume} {88}},\
  \bibinfo {pages} {121406} (\bibinfo {year} {2013})}\BibitemShut {NoStop}%
\bibitem [{\citenamefont {D'Alessio}\ and\ \citenamefont
  {Rigol}(2014)}]{DAlessio14}%
  \BibitemOpen
  \bibfield  {author} {\bibinfo {author} {\bibfnamefont {L.}~\bibnamefont
  {D'Alessio}}\ and\ \bibinfo {author} {\bibfnamefont {M.}~\bibnamefont
  {Rigol}},\ }\href {\doibase 10.1103/PhysRevX.4.041048} {\bibfield  {journal}
  {\bibinfo  {journal} {Phys. Rev. X}\ }\textbf {\bibinfo {volume} {4}},\
  \bibinfo {pages} {041048} (\bibinfo {year} {2014})}\BibitemShut {NoStop}%
\bibitem [{\citenamefont {Lazarides}\ \emph {et~al.}(2014)\citenamefont
  {Lazarides}, \citenamefont {Das},\ and\ \citenamefont
  {Moessner}}]{Lazarides14}%
  \BibitemOpen
  \bibfield  {author} {\bibinfo {author} {\bibfnamefont {A.}~\bibnamefont
  {Lazarides}}, \bibinfo {author} {\bibfnamefont {A.}~\bibnamefont {Das}}, \
  and\ \bibinfo {author} {\bibfnamefont {R.}~\bibnamefont {Moessner}},\ }\href
  {\doibase 10.1103/PhysRevE.90.012110} {\bibfield  {journal} {\bibinfo
  {journal} {Phys. Rev. E}\ }\textbf {\bibinfo {volume} {90}},\ \bibinfo
  {pages} {012110} (\bibinfo {year} {2014})}\BibitemShut {NoStop}%
\bibitem [{\citenamefont {Ponte}\ \emph
  {et~al.}(2015{\natexlab{a}})\citenamefont {Ponte}, \citenamefont {Chandran},
  \citenamefont {PapiÄ},\ and\ \citenamefont {Abanin}}]{Ponte15}%
  \BibitemOpen
  \bibfield  {author} {\bibinfo {author} {\bibfnamefont {P.}~\bibnamefont
  {Ponte}}, \bibinfo {author} {\bibfnamefont {A.}~\bibnamefont {Chandran}},
  \bibinfo {author} {\bibfnamefont {Z.}~\bibnamefont {PapiÄ}}, \ and\
  \bibinfo {author} {\bibfnamefont {D.~A.}\ \bibnamefont {Abanin}},\ }\href
  {\doibase http://dx.doi.org/10.1016/j.aop.2014.11.008} {\bibfield  {journal}
  {\bibinfo  {journal} {Annals of Physics}\ }\textbf {\bibinfo {volume}
  {353}},\ \bibinfo {pages} {196 } (\bibinfo {year}
  {2015}{\natexlab{a}})}\BibitemShut {NoStop}%
\bibitem [{\citenamefont {{Khemani}}\ \emph {et~al.}(2015)\citenamefont
  {{Khemani}}, \citenamefont {{Lazarides}}, \citenamefont {{Moessner}},\ and\
  \citenamefont {{Sondhi}}}]{Khemani15}%
  \BibitemOpen
  \bibfield  {author} {\bibinfo {author} {\bibfnamefont {V.}~\bibnamefont
  {{Khemani}}}, \bibinfo {author} {\bibfnamefont {A.}~\bibnamefont
  {{Lazarides}}}, \bibinfo {author} {\bibfnamefont {R.}~\bibnamefont
  {{Moessner}}}, \ and\ \bibinfo {author} {\bibfnamefont {S.~L.}\ \bibnamefont
  {{Sondhi}}},\ }\href@noop {} {\bibfield  {journal} {\bibinfo  {journal}
  {ArXiv e-prints}\ } (\bibinfo {year} {2015})},\ \Eprint
  {http://arxiv.org/abs/1508.03344} {arXiv:1508.03344 [cond-mat.dis-nn]}
  \BibitemShut {NoStop}%
\bibitem [{\citenamefont {Basko}\ \emph {et~al.}(2006)\citenamefont {Basko},
  \citenamefont {Aleiner},\ and\ \citenamefont {Altshuler}}]{Basko06}%
  \BibitemOpen
  \bibfield  {author} {\bibinfo {author} {\bibfnamefont {D.}~\bibnamefont
  {Basko}}, \bibinfo {author} {\bibfnamefont {I.}~\bibnamefont {Aleiner}}, \
  and\ \bibinfo {author} {\bibfnamefont {B.}~\bibnamefont {Altshuler}},\ }\href
  {\doibase http://dx.doi.org/10.1016/j.aop.2005.11.014} {\bibfield  {journal}
  {\bibinfo  {journal} {Annals of Physics}\ }\textbf {\bibinfo {volume}
  {321}},\ \bibinfo {pages} {1126 } (\bibinfo {year} {2006})}\BibitemShut
  {NoStop}%
\bibitem [{\citenamefont {Nandkishore}\ and\ \citenamefont
  {Huse}(2015)}]{Nandkishore15}%
  \BibitemOpen
  \bibfield  {author} {\bibinfo {author} {\bibfnamefont {R.}~\bibnamefont
  {Nandkishore}}\ and\ \bibinfo {author} {\bibfnamefont {D.~A.}\ \bibnamefont
  {Huse}},\ }\href {\doibase 10.1146/annurev-conmatphys-031214-014726}
  {\bibfield  {journal} {\bibinfo  {journal} {Annual Review of Condensed Matter
  Physics}\ }\textbf {\bibinfo {volume} {6}},\ \bibinfo {pages} {15} (\bibinfo
  {year} {2015})},\ \Eprint
  {http://arxiv.org/abs/http://dx.doi.org/10.1146/annurev-conmatphys-031214-014726}
  {http://dx.doi.org/10.1146/annurev-conmatphys-031214-014726} \BibitemShut
  {NoStop}%
\bibitem [{\citenamefont {Oganesyan}\ and\ \citenamefont
  {Huse}(2007)}]{Oganesyan07}%
  \BibitemOpen
  \bibfield  {author} {\bibinfo {author} {\bibfnamefont {V.}~\bibnamefont
  {Oganesyan}}\ and\ \bibinfo {author} {\bibfnamefont {D.~A.}\ \bibnamefont
  {Huse}},\ }\href {\doibase 10.1103/PhysRevB.75.155111} {\bibfield  {journal}
  {\bibinfo  {journal} {Phys. Rev. B}\ }\textbf {\bibinfo {volume} {75}},\
  \bibinfo {pages} {155111} (\bibinfo {year} {2007})}\BibitemShut {NoStop}%
\bibitem [{\citenamefont {Pal}\ and\ \citenamefont {Huse}(2010)}]{Pal10}%
  \BibitemOpen
  \bibfield  {author} {\bibinfo {author} {\bibfnamefont {A.}~\bibnamefont
  {Pal}}\ and\ \bibinfo {author} {\bibfnamefont {D.~A.}\ \bibnamefont {Huse}},\
  }\href {\doibase 10.1103/PhysRevB.82.174411} {\bibfield  {journal} {\bibinfo
  {journal} {Phys. Rev. B}\ }\textbf {\bibinfo {volume} {82}},\ \bibinfo
  {pages} {174411} (\bibinfo {year} {2010})}\BibitemShut {NoStop}%
\bibitem [{\citenamefont {{Abanin}}\ \emph {et~al.}(2014)\citenamefont
  {{Abanin}}, \citenamefont {{De Roeck}},\ and\ \citenamefont
  {{Huveneers}}}]{Abanin14}%
  \BibitemOpen
  \bibfield  {author} {\bibinfo {author} {\bibfnamefont {D.}~\bibnamefont
  {{Abanin}}}, \bibinfo {author} {\bibfnamefont {W.}~\bibnamefont {{De
  Roeck}}}, \ and\ \bibinfo {author} {\bibfnamefont {F.}~\bibnamefont
  {{Huveneers}}},\ }\href@noop {} {\bibfield  {journal} {\bibinfo  {journal}
  {ArXiv e-prints}\ } (\bibinfo {year} {2014})},\ \Eprint
  {http://arxiv.org/abs/1412.4752} {arXiv:1412.4752 [cond-mat.dis-nn]}
  \BibitemShut {NoStop}%
\bibitem [{\citenamefont {Ponte}\ \emph
  {et~al.}(2015{\natexlab{b}})\citenamefont {Ponte}, \citenamefont
  {Papi\ifmmode~\acute{c}\else \'{c}\fi{}}, \citenamefont {Huveneers},\ and\
  \citenamefont {Abanin}}]{Ponte14}%
  \BibitemOpen
  \bibfield  {author} {\bibinfo {author} {\bibfnamefont {P.}~\bibnamefont
  {Ponte}}, \bibinfo {author} {\bibfnamefont {Z.}~\bibnamefont
  {Papi\ifmmode~\acute{c}\else \'{c}\fi{}}}, \bibinfo {author} {\bibfnamefont
  {F.~m.~c.}\ \bibnamefont {Huveneers}}, \ and\ \bibinfo {author}
  {\bibfnamefont {D.~A.}\ \bibnamefont {Abanin}},\ }\href {\doibase
  10.1103/PhysRevLett.114.140401} {\bibfield  {journal} {\bibinfo  {journal}
  {Phys. Rev. Lett.}\ }\textbf {\bibinfo {volume} {114}},\ \bibinfo {pages}
  {140401} (\bibinfo {year} {2015}{\natexlab{b}})}\BibitemShut {NoStop}%
\bibitem [{\citenamefont {Lazarides}\ \emph {et~al.}(2015)\citenamefont
  {Lazarides}, \citenamefont {Das},\ and\ \citenamefont
  {Moessner}}]{Lazarides15}%
  \BibitemOpen
  \bibfield  {author} {\bibinfo {author} {\bibfnamefont {A.}~\bibnamefont
  {Lazarides}}, \bibinfo {author} {\bibfnamefont {A.}~\bibnamefont {Das}}, \
  and\ \bibinfo {author} {\bibfnamefont {R.}~\bibnamefont {Moessner}},\ }\href
  {\doibase 10.1103/PhysRevLett.115.030402} {\bibfield  {journal} {\bibinfo
  {journal} {Phys. Rev. Lett.}\ }\textbf {\bibinfo {volume} {115}},\ \bibinfo
  {pages} {030402} (\bibinfo {year} {2015})}\BibitemShut {NoStop}%
\bibitem [{\citenamefont {Huse}\ \emph {et~al.}(2013)\citenamefont {Huse},
  \citenamefont {Nandkishore}, \citenamefont {Oganesyan}, \citenamefont {Pal},\
  and\ \citenamefont {Sondhi}}]{Huse13}%
  \BibitemOpen
  \bibfield  {author} {\bibinfo {author} {\bibfnamefont {D.~A.}\ \bibnamefont
  {Huse}}, \bibinfo {author} {\bibfnamefont {R.}~\bibnamefont {Nandkishore}},
  \bibinfo {author} {\bibfnamefont {V.}~\bibnamefont {Oganesyan}}, \bibinfo
  {author} {\bibfnamefont {A.}~\bibnamefont {Pal}}, \ and\ \bibinfo {author}
  {\bibfnamefont {S.~L.}\ \bibnamefont {Sondhi}},\ }\href {\doibase
  10.1103/PhysRevB.88.014206} {\bibfield  {journal} {\bibinfo  {journal} {Phys.
  Rev. B}\ }\textbf {\bibinfo {volume} {88}},\ \bibinfo {pages} {014206}
  (\bibinfo {year} {2013})}\BibitemShut {NoStop}%
\bibitem [{\citenamefont {Pekker}\ \emph {et~al.}(2014)\citenamefont {Pekker},
  \citenamefont {Refael}, \citenamefont {Altman}, \citenamefont {Demler},\ and\
  \citenamefont {Oganesyan}}]{Pekker14}%
  \BibitemOpen
  \bibfield  {author} {\bibinfo {author} {\bibfnamefont {D.}~\bibnamefont
  {Pekker}}, \bibinfo {author} {\bibfnamefont {G.}~\bibnamefont {Refael}},
  \bibinfo {author} {\bibfnamefont {E.}~\bibnamefont {Altman}}, \bibinfo
  {author} {\bibfnamefont {E.}~\bibnamefont {Demler}}, \ and\ \bibinfo {author}
  {\bibfnamefont {V.}~\bibnamefont {Oganesyan}},\ }\href {\doibase
  10.1103/PhysRevX.4.011052} {\bibfield  {journal} {\bibinfo  {journal} {Phys.
  Rev. X}\ }\textbf {\bibinfo {volume} {4}},\ \bibinfo {pages} {011052}
  (\bibinfo {year} {2014})}\BibitemShut {NoStop}%
\bibitem [{\citenamefont {Chandran}\ \emph {et~al.}(2014)\citenamefont
  {Chandran}, \citenamefont {Khemani}, \citenamefont {Laumann},\ and\
  \citenamefont {Sondhi}}]{Chandran14}%
  \BibitemOpen
  \bibfield  {author} {\bibinfo {author} {\bibfnamefont {A.}~\bibnamefont
  {Chandran}}, \bibinfo {author} {\bibfnamefont {V.}~\bibnamefont {Khemani}},
  \bibinfo {author} {\bibfnamefont {C.~R.}\ \bibnamefont {Laumann}}, \ and\
  \bibinfo {author} {\bibfnamefont {S.~L.}\ \bibnamefont {Sondhi}},\ }\href
  {\doibase 10.1103/PhysRevB.89.144201} {\bibfield  {journal} {\bibinfo
  {journal} {Phys. Rev. B}\ }\textbf {\bibinfo {volume} {89}},\ \bibinfo
  {pages} {144201} (\bibinfo {year} {2014})}\BibitemShut {NoStop}%
\bibitem [{\citenamefont {Kj\"all}\ \emph {et~al.}(2014)\citenamefont
  {Kj\"all}, \citenamefont {Bardarson},\ and\ \citenamefont
  {Pollmann}}]{Kjall14}%
  \BibitemOpen
  \bibfield  {author} {\bibinfo {author} {\bibfnamefont {J.~A.}\ \bibnamefont
  {Kj\"all}}, \bibinfo {author} {\bibfnamefont {J.~H.}\ \bibnamefont
  {Bardarson}}, \ and\ \bibinfo {author} {\bibfnamefont {F.}~\bibnamefont
  {Pollmann}},\ }\href {\doibase 10.1103/PhysRevLett.113.107204} {\bibfield
  {journal} {\bibinfo  {journal} {Phys. Rev. Lett.}\ }\textbf {\bibinfo
  {volume} {113}},\ \bibinfo {pages} {107204} (\bibinfo {year}
  {2014})}\BibitemShut {NoStop}%
\bibitem [{\citenamefont {Vosk}\ and\ \citenamefont {Altman}(2014)}]{Vosk14}%
  \BibitemOpen
  \bibfield  {author} {\bibinfo {author} {\bibfnamefont {R.}~\bibnamefont
  {Vosk}}\ and\ \bibinfo {author} {\bibfnamefont {E.}~\bibnamefont {Altman}},\
  }\href {\doibase 10.1103/PhysRevLett.112.217204} {\bibfield  {journal}
  {\bibinfo  {journal} {Phys. Rev. Lett.}\ }\textbf {\bibinfo {volume} {112}},\
  \bibinfo {pages} {217204} (\bibinfo {year} {2014})}\BibitemShut {NoStop}%
\bibitem [{\citenamefont {Bahri}\ \emph {et~al.}(2015)\citenamefont {Bahri},
  \citenamefont {Vosk}, \citenamefont {Altman},\ and\ \citenamefont
  {Vishwanath}}]{Bahri15}%
  \BibitemOpen
  \bibfield  {author} {\bibinfo {author} {\bibfnamefont {Y.}~\bibnamefont
  {Bahri}}, \bibinfo {author} {\bibfnamefont {R.}~\bibnamefont {Vosk}},
  \bibinfo {author} {\bibfnamefont {E.}~\bibnamefont {Altman}}, \ and\ \bibinfo
  {author} {\bibfnamefont {A.}~\bibnamefont {Vishwanath}},\ }\href
  {http://dx.doi.org/10.1038/ncomms8341} {\bibfield  {journal} {\bibinfo
  {journal} {Nat Commun}\ }\textbf {\bibinfo {volume} {6}} (\bibinfo {year}
  {2015})}\BibitemShut {NoStop}%
\bibitem [{\citenamefont {Mondragon-Shem}\ \emph {et~al.}(2014)\citenamefont
  {Mondragon-Shem}, \citenamefont {Hughes}, \citenamefont {Song},\ and\
  \citenamefont {Prodan}}]{Mondragon-Shem14}%
  \BibitemOpen
  \bibfield  {author} {\bibinfo {author} {\bibfnamefont {I.}~\bibnamefont
  {Mondragon-Shem}}, \bibinfo {author} {\bibfnamefont {T.~L.}\ \bibnamefont
  {Hughes}}, \bibinfo {author} {\bibfnamefont {J.}~\bibnamefont {Song}}, \ and\
  \bibinfo {author} {\bibfnamefont {E.}~\bibnamefont {Prodan}},\ }\href
  {\doibase 10.1103/PhysRevLett.113.046802} {\bibfield  {journal} {\bibinfo
  {journal} {Phys. Rev. Lett.}\ }\textbf {\bibinfo {volume} {113}},\ \bibinfo
  {pages} {046802} (\bibinfo {year} {2014})}\BibitemShut {NoStop}%
\bibitem [{\citenamefont {Song}\ and\ \citenamefont {Prodan}(2014)}]{Song14}%
  \BibitemOpen
  \bibfield  {author} {\bibinfo {author} {\bibfnamefont {J.}~\bibnamefont
  {Song}}\ and\ \bibinfo {author} {\bibfnamefont {E.}~\bibnamefont {Prodan}},\
  }\href {\doibase 10.1103/PhysRevB.89.224203} {\bibfield  {journal} {\bibinfo
  {journal} {Phys. Rev. B}\ }\textbf {\bibinfo {volume} {89}},\ \bibinfo
  {pages} {224203} (\bibinfo {year} {2014})}\BibitemShut {NoStop}%
\bibitem [{\citenamefont {Tarasinski}\ \emph {et~al.}(2014)\citenamefont
  {Tarasinski}, \citenamefont {Asb\'oth},\ and\ \citenamefont
  {Dahlhaus}}]{Tarasinski14}%
  \BibitemOpen
  \bibfield  {author} {\bibinfo {author} {\bibfnamefont {B.}~\bibnamefont
  {Tarasinski}}, \bibinfo {author} {\bibfnamefont {J.~K.}\ \bibnamefont
  {Asb\'oth}}, \ and\ \bibinfo {author} {\bibfnamefont {J.~P.}\ \bibnamefont
  {Dahlhaus}},\ }\href {\doibase 10.1103/PhysRevA.89.042327} {\bibfield
  {journal} {\bibinfo  {journal} {Phys. Rev. A}\ }\textbf {\bibinfo {volume}
  {89}},\ \bibinfo {pages} {042327} (\bibinfo {year} {2014})}\BibitemShut
  {NoStop}%
\bibitem [{\citenamefont {Obuse}\ and\ \citenamefont
  {Kawakami}(2011)}]{Obuse11}%
  \BibitemOpen
  \bibfield  {author} {\bibinfo {author} {\bibfnamefont {H.}~\bibnamefont
  {Obuse}}\ and\ \bibinfo {author} {\bibfnamefont {N.}~\bibnamefont
  {Kawakami}},\ }\href {\doibase 10.1103/PhysRevB.84.195139} {\bibfield
  {journal} {\bibinfo  {journal} {Phys. Rev. B}\ }\textbf {\bibinfo {volume}
  {84}},\ \bibinfo {pages} {195139} (\bibinfo {year} {2011})}\BibitemShut
  {NoStop}%
\bibitem [{\citenamefont {Ryu}\ \emph {et~al.}(2010)\citenamefont {Ryu},
  \citenamefont {Schnyder}, \citenamefont {Furusaki},\ and\ \citenamefont
  {Ludwig}}]{Ryu10}%
  \BibitemOpen
  \bibfield  {author} {\bibinfo {author} {\bibfnamefont {S.}~\bibnamefont
  {Ryu}}, \bibinfo {author} {\bibfnamefont {A.~P.}\ \bibnamefont {Schnyder}},
  \bibinfo {author} {\bibfnamefont {A.}~\bibnamefont {Furusaki}}, \ and\
  \bibinfo {author} {\bibfnamefont {A.~W.~W.}\ \bibnamefont {Ludwig}},\ }\href
  {http://stacks.iop.org/1367-2630/12/i=6/a=065010} {\bibfield  {journal}
  {\bibinfo  {journal} {New Journal of Physics}\ }\textbf {\bibinfo {volume}
  {12}},\ \bibinfo {pages} {065010} (\bibinfo {year} {2010})}\BibitemShut
  {NoStop}%
\bibitem [{\citenamefont {Kramer}\ and\ \citenamefont
  {MacKinnon}(1993)}]{Kramer93}%
  \BibitemOpen
  \bibfield  {author} {\bibinfo {author} {\bibfnamefont {B.}~\bibnamefont
  {Kramer}}\ and\ \bibinfo {author} {\bibfnamefont {A.}~\bibnamefont
  {MacKinnon}},\ }\href {http://stacks.iop.org/0034-4885/56/i=12/a=001}
  {\bibfield  {journal} {\bibinfo  {journal} {Reports on Progress in Physics}\
  }\textbf {\bibinfo {volume} {56}},\ \bibinfo {pages} {1469} (\bibinfo {year}
  {1993})}\BibitemShut {NoStop}%
\bibitem [{\citenamefont {Bukov}\ \emph {et~al.}(2015)\citenamefont {Bukov},
  \citenamefont {D'Alessio},\ and\ \citenamefont {Polkovnikov}}]{Bukov15}%
  \BibitemOpen
  \bibfield  {author} {\bibinfo {author} {\bibfnamefont {M.}~\bibnamefont
  {Bukov}}, \bibinfo {author} {\bibfnamefont {L.}~\bibnamefont {D'Alessio}}, \
  and\ \bibinfo {author} {\bibfnamefont {A.}~\bibnamefont {Polkovnikov}},\
  }\href {\doibase 10.1080/00018732.2015.1055918} {\bibfield  {journal}
  {\bibinfo  {journal} {Advances in Physics}\ }\textbf {\bibinfo {volume}
  {64}},\ \bibinfo {pages} {139} (\bibinfo {year} {2015})},\ \Eprint
  {http://arxiv.org/abs/http://dx.doi.org/10.1080/00018732.2015.1055918}
  {http://dx.doi.org/10.1080/00018732.2015.1055918} \BibitemShut {NoStop}%
\bibitem [{\citenamefont {Rakovszky}\ and\ \citenamefont
  {Asboth}(2015)}]{Rakovszky15}%
  \BibitemOpen
  \bibfield  {author} {\bibinfo {author} {\bibfnamefont {T.}~\bibnamefont
  {Rakovszky}}\ and\ \bibinfo {author} {\bibfnamefont {J.~K.}\ \bibnamefont
  {Asboth}},\ }\href {\doibase 10.1103/PhysRevA.92.052311} {\bibfield
  {journal} {\bibinfo  {journal} {Phys. Rev. A}\ }\textbf {\bibinfo {volume}
  {92}},\ \bibinfo {pages} {052311} (\bibinfo {year} {2015})}\BibitemShut
  {NoStop}%
\end{thebibliography}%

\end{document}